\documentstyle[aps]{revtex}
\begin{document}
\title{{\bf Consistency in Perturbative Calculations and Radiatively Induced
Lorentz and CPT Violations}}
\author{O.A. Battistel* and G. Dallabona**}
\maketitle

\centerline{*Dept. of Physics-CCNE, Universidade Federal de Santa Maria}

\centerline{P.O. Box 5093, 97119-900, Santa Maria, RS, Brazil}

\centerline{orimar@ccne.ufsm.br}

\centerline{**Dept. of Physics-ICEx, Universidade Federal de Minas Gerais}

\centerline{P.O. Box 702, 30161-970, Belo Horizonte, MG, Brazil}

\centerline{dalla@fisica.ufmg.br}

\begin{abstract}
The origin of the radiatively induced Lorentz and CPT violations, in
perturbative evaluations, of an extended version of QED, is investigated.
Using a very general calculational method, concerning the manipulations and
calculations involving divergent amplitudes, we clearly identify the
possible sources of contributions for the violating terms. We show that
consistency in the perturbative calculations, in a broader sense, leaves no
room for the existence of radiatively induced contributions which is in
accordance with what was previously conjectured and recently advocated by
some authors supported on general arguments.
\end{abstract}

\vskip 0.5cm

\noindent PACS numbers: 11.15.Bt; 11.30.Qc; 11.30.Cp; 11.30.Er

\noindent The implications of Lorentz and CPT symmetry breaking have
received a great deal of attention in the last few years. Most of them were
all dedicated to the QED extended sector of the Extended Standard Model
constructed by Colladay and Kostelecky \cite{Colladay-Kostelecky1}. They
have developed a conceptual framework and a procedure for the treatment of
spontaneous CPT and Lorentz violation within a context where the gauge
structure and renormalizability are maintained \cite{Colladay-Kostelecky2}.
The Lagrangian of the QED extended sector is composed by the usual QED
theory in addition to the breaking terms 
\begin{equation}
L^{SB}=-a_{\mu }\bar{\Psi}\gamma ^{\mu }\Psi -b_{\mu }\bar{\Psi}\gamma
_{5}\gamma ^{\mu }\Psi +\frac{1}{2}k^{\alpha }\epsilon _{\alpha \lambda \mu
\nu }A^{\lambda }F^{\mu \nu }.
\end{equation}
In the above expression, $a_{\mu }$ and $b_{\mu }$ are real and constant
prescribed four-vectors. The coupling $k_{\alpha }$ is also real and it has
dimension of mass. The matrix $\gamma _{5}$ is the usual Hermitian Dirac
matrix which is related to the totally antisymmetric tensor $\varepsilon
_{\mu \nu \alpha \beta }$ through $tr{\gamma _{5}\gamma _{\alpha }\gamma
_{\beta }\gamma _{\mu }\gamma _{\nu }}=4i\epsilon _{\alpha \beta \mu \nu }$.
The mathematical structure of the purely photonic sector breaking term, in
the above Lagrangian, allows us to identify an important consequence for the
modified QED theory. Due to the fact that it changes by a total derivative
under potential gauge transformation $(A_{\mu }\rightarrow A_{\mu }+\partial
_{\mu }\Lambda ),$ the action is not modified and the resulting equations of
motions remain the same ones as those of the original theory. Such behavior
is precisely what we call the Chern-Simons(CS) form. In spite of this, there
are many phenomenological consequences associated to the modified theory 
\cite{Colladay-Kostelecky1} -\cite{Coleman-Glashow}. However, all the
present experimental and theoretical investigations seem to state that the $%
k_{\mu }$ coupling value, compatible with the phenomenology, is the
identically zero one. Such statement does not completely eliminate the
possibility of Lorentz and CPT breaking effects being present in the
modified theory. Even that the $k_{\mu }$ coupling vanishes at the tree
level, such type of effects can be, in principle, induced by radiative
contributions. In the QED extended theory, radiative corrections coming from
the fermionic sector can induce contributions of the CS form \cite
{Colladay-Kostelecky2}. It could appear when the photon propagator is
corrected by the $b$-breaking term. From the calculational point of view, we
have to evaluate the usual QED one-loop vacuum polarization tensor with the
free spin-$1/2$ fermion propagator, obeying the Dirac equation, changed by
the inclusion of the $b_{\mu }$ coupling: 
\begin{equation}
G(k)=\frac{i}{\not{k}-m-\not{b}\gamma _{5}}.
\end{equation}
The corresponding one-loop amplitude can be written as 
\begin{equation}
\Pi ^{\mu \nu }(p)=\int \frac{d^{4}k}{(2\pi )^{4}}tr\left\{ \gamma ^{\mu
}G(k)\gamma ^{\nu }G(k+p)\right\} .
\end{equation}
The evaluation of the above amplitude has been performed by many authors and
different aspects involved in the calculations were emphasized \cite
{Colladay-Kostelecky2}\cite{Jackiw1} -\cite{Kostelecky3} \cite
{Coleman-Glashow}\cite{Bonneau} -\cite{Orimar1-2}. For a first set of
authors and works, the obtained result is nonzero but it is essentially
ambiguous. As a consequence, a definite value can be only stated after an
arbitrary set of choices is taken. So, the results presented by the referred
authors represent only a particular choice for the involved ambiguities.
More recently, a second set of authors have argued, by using very general
arguments, that only the definite zero value for the radiatively induced CS
term is reasonable. Among them, G. Bonneau \cite{Bonneau} shows that, if the
theory is correctly defined by taking into account their Ward identities and
appropriated normalization conditions, the CS term is absent in a
non-ambiguous way. On the other hand, C. Adam and F.R. Klinkhamer \cite{Adam}%
, by requiring causality in addition to the validity of the perturbation
theory, concluded that no CS term must exist. The same result was previously
conjectured by Colladay and Kostelecky \cite{Colladay-Kostelecky2} and
advocated by Coleman and Glashow \cite{Coleman-Glashow}. So, it remains a
question: Why do general physical arguments lead to a definite zero value
while the perturbative calculations do not? On the other hand, given the
fact that a nonzero Chern-Simons term represents Lorentz and CPT violation,
we can ask which are the steps in the perturbative evaluation that give
raise to the non-zero contribution. Talking in different words, which are
the assumptions, taken in the intermediate steps, where resides the origin
of the Lorentz and CPT violation in the perturbative calculation, that a
consistent handling of divergences may avoid in order to obtain the expected
value for the CS term, the definite zero value?

The main point is the following: If the CS term is non-vanishing, it
represents a Lorentz and CPT symmetry breaking induced by radiative
corrections in the QED extended theory. However, in order to consider such
phenomenon as a fundamental one, it must not be a simple consequence of a
choice for arbitrariness, but it should emerge as an unavoidable aspect of
the calculations. In other words, to define a calculational scheme such that
some symmetries are violated in the calculations is a very simple job, but
this does not mean that the corresponding phenomenon must exist. We have to
put this particular calculation in accordance with other ones of the
perturbative calculations, specially those which are necessary for the
construction of the (symmetric) renormalizable Standard Model. This means to
treat all the involved mathematical structures in the same way they are
treated in the symmetric theory.

The purpose of the present work is precisely to clarify these points. We
will use a very general calculational strategy to handle divergences \cite
{Orimar2} in order to isolate in a very clear way the possible contributions
for the CS term in the perturbative evaluation, and show that when the
interpretation required by the consistency in perturbative calculations is
adopted, in a broader sense, an exactly zero value for the Lorentz and CPT
contribution is achieved.

In order to evaluate the CS term we have to consider $\gamma _{5}$-odd
divergent structures and therefore the Dimensional Regularization (DR)
technique \cite{DR} is excluded from the possible tools. To make clear this
point we write in the expression (3) the exact propagators $G(k)$, given in
the expression (2), in the form \cite{Jackiw1} 
\begin{equation}
G(l)=S(l)+G_{b}(l),
\end{equation}
with 
\begin{equation}
G_{b}(l)=\frac{1}{\not{k}-m-\not{b}\gamma _{5}}\not{b}\gamma _{5}S(l),
\end{equation}
and $S(l)\ $being the usual spin-$1/2$ fermion propagator. After the
substitution of the above expression, the $\Pi _{\mu \nu }(p)$ amplitude can
be split in three terms: $\Pi ^{\mu \nu }=\Pi _{0}^{\mu \nu }+\Pi _{b}^{\mu
\nu }+\Pi _{bb}^{\mu \nu }$. The first contribution, $\Pi _{0}^{\mu \nu },$
is precisely the pure QED vacuum polarization tensor. The linear b-term is
given by 
\begin{equation}
\Pi _{b}^{\mu \nu }(p)=\int \frac{d^{4}k}{(2\pi )^{4}}tr\left\{ \gamma ^{\mu
}S(l)\gamma ^{\nu }G_{b}(l+p)+\gamma ^{\mu }G_{b}(l+p)\gamma ^{\nu
}S(l)\right\} .
\end{equation}
To evaluate $\Pi _{b}^{\mu \nu }$ to the lowest order in $b$, we simply
replace the expression (5) by 
\begin{equation}
G_{b}(k)=-iS(k)\not{b}\gamma _{5}S(k).
\end{equation}
Now, the corresponding expression to $\Pi _{b}^{\mu \nu }$ may be written as 
\begin{equation}
\Pi _{b}^{\mu \nu }(p)\simeq b_{\lambda }\Pi ^{\mu \nu \lambda }(p),
\end{equation}
where 
\begin{eqnarray}
\Pi ^{\mu \nu \lambda }(p) &=&(-i)\int \frac{d^{4}k}{(2\pi )^{4}}tr\left\{
\gamma ^{\mu }S(k)\gamma ^{\nu }S(k+p)\gamma ^{\lambda }\gamma
_{5}S(k+p)\right. +  \nonumber \\
&&\;\;\;\;\;\;\;\;\;\;\;\;\;\;\;\;\;\;\;\;\;\;\;\;\;+\left. \gamma ^{\mu
}S(k)\gamma ^{\lambda }\gamma _{5}S(k)\gamma ^{\nu }S(k+p)\right\} .
\end{eqnarray}
So, the crucial mathematical structure, which we need to evaluate in order
to get the value for the CS term, is an $AVV$ triangle amplitude. Such
three-point function is a Green's function of the symmetric theory (the
renormalizable Standard Model). There are many kinds of arbitrariness
involved in the evaluation of such amplitude. The requirement of consistency
in the perturbative calculation implies that the choices for the
arbitrariness present in the above expression must be taken in a consistent
way with those adopted in the construction of the symmetric Standard Model.
Given this argument we will consider the most general mathematical
expression and only after all the considerations relative to the consistency
in the evaluation of perturbative amplitudes have been made, we will return
to the specific situation of the eq.(9).

We start by the definition 
\begin{equation}
T_{\lambda \mu \nu }^{AVV}=\int \frac{d^{4}k}{\left( 2\pi \right) ^{4}}%
tr\left\{ \gamma _{\mu }\left[ (\not{k}+\not{k}_{1})-m\right] ^{-1}\gamma
_{\nu }\left[ (\not{k}+\not{k}_{2})-m\right] ^{-1}i\gamma _{\lambda }\gamma
_{5}\left[ (\not{k}+\not{k}_{3})-m\right] ^{-1}\right\} ,
\end{equation}
which corresponds to the most general expression for the direct diagram. In
the above expression $k_{1},\;k_{2}$ and $k_{3}$ stand for arbitrary choices
for the internal lines momenta. They are related to the external ones by
their differences which we adopt: $k_{3}-k_{1}=p,k_{1}-k_{2}=p^{\prime }$
and $k_{3}-k_{2}=p^{\prime }+p=q$. Their summations are undefined physical
quantities. If we are worried about consistency, in the evaluation of this
superficially linearly divergent structure, the first step is the
identification of eventual constraints that this amplitude should obey, in
spite of its divergent character. Such constraints are invariably
materialized through relations among other amplitudes and/or by fixing a
kinematical limit through a low-energy theorem. Due to the fact that, in
principle, all the Green's functions of the perturbative expansion are in
some way related, we can use eventual constraints imposed by general
physical grounds upon a particular amplitude to restrict other ones. For the 
$AVV$ we note, for example, the identity 
\begin{eqnarray}
&&(k_{3}-k_{2})_{\lambda }\left\{ \gamma _{\nu }\frac{1}{(\not{k}+{\not{k}}%
_{2})-m}i\gamma _{\lambda }\gamma _{5}\frac{1}{(\not{k}+{\not{k}}_{3})-m}%
\gamma _{\mu }\frac{1}{(\not{k}+{\not{k}}_{1})-m}\right\} =-\left\{ i\gamma
_{\nu }\gamma _{5}\frac{1}{(\not{k}+{\not{k}}_{3})-m}\gamma _{\mu }\frac{1}{(%
\not{k}+{\not{k}}_{1})-m}\right\}  \nonumber \\
&&-2mi\left\{ \gamma _{\nu }\frac{1}{(\not{k}+{\not{k}}_{2})-m}\gamma _{5}%
\frac{1}{(\not{k}+{\not{k}}_{3})-m}\gamma _{\mu }\frac{1}{(\not{k}+{\not{k}}%
_{1})-m}\right\} +\left\{ \gamma _{\nu }\frac{1}{(\not{k}+{\not{k}}_{2})-m}%
i\gamma _{\mu }\gamma _{5}\frac{1}{(\not{k}+{\not{k}}_{1})-m}\right\} .
\end{eqnarray}
The above identity, which has nothing to do with divergences, can be
converted to a relation among perturbative Green's functions if we take the
traces operation in both sides and next integrate in the momentum $k$. This
gives us 
\begin{equation}
\left( k_{3}-k_{2}\right) _{\lambda }T_{\lambda \mu \nu }^{AVV}\left(
k_{1},k_{2},k_{3};m\right) =-2miT_{\mu \nu }^{PVV}\left(
k_{1},k_{2},k_{3};m\right) +T_{\mu \nu }^{AV}\left( k_{1},k_{2};m\right)
-T_{\nu \mu }^{AV}\left( k_{3},k_{1};m\right) ,
\end{equation}
where we have introduced the $PVV$ three-point function defined by 
\begin{equation}
T_{\mu \nu }^{PVV}\left( k_{1},k_{2},k_{3};m\right) =\int \frac{d^{4}k}{%
(2\pi )^{4}}Tr\left\{ \gamma _{5}\frac{1}{(\not{k}+{\not{k}}_{3})-m}\gamma
_{\mu }\frac{1}{(\not{k}+{\not{k}}_{1})-m}\gamma _{\nu }\frac{1}{(\not{k}+{%
\not{k}}_{2})-m}\right\} ,
\end{equation}
and the $AV$ two-point function 
\begin{equation}
T_{\mu \nu }^{AV}\left( k_{1},k_{2};m\right) =\int \frac{d^{4}k}{(2\pi )^{4}}%
Tr\left\{ i\gamma _{\mu }\gamma _{5}\frac{1}{(\not{k}+{\not{k}}_{1})-m}%
\gamma _{\nu }\frac{1}{(\not{k}+{\not{k}}_{2})-m}\right\} .
\end{equation}
Following a similar procedure, two other relations can be produced by
contracting the term between curly brackets on the left hand side of the
eq.(11) with the external momenta $\left( k_{3}-k_{1}\right) _{\mu }$ and $%
\left( k_{1}-k_{2}\right) _{\nu }$. They are 
\begin{eqnarray}
\bullet (k_{3}-k_{1})_{\mu }T_{{\lambda }{\mu }{\nu }}^{AVV} &=&T_{{\lambda
\nu }}^{AV}(k_{1},k_{2};m)-T_{{\lambda \nu }}^{AV}(k_{3},k_{2};m) \\
\bullet (k_{1}-k_{2})_{\nu }T_{{\lambda }{\mu }{\nu }}^{AVV} &=&T_{{\lambda
\mu }}^{AV}(k_{3},k_{2};m)-T_{{\lambda \mu }}^{AV}(k_{3},k_{1};m).
\end{eqnarray}
The $AV$ structure that appeared on the right hand side of the eqs.(12),
(15) and (16), is a two-point function physical amplitude and possesses its
own relations with other physical amplitudes. The most important one for our
present purposes is the following 
\begin{equation}
T_{\mu \nu }^{AV}(k_{1},k_{2};m)=-\frac{1}{2m}\varepsilon _{\mu \nu \alpha
\beta }(k_{1}-k_{2})_{\alpha }T_{\beta }^{SV}(k_{1},k_{2};m),
\end{equation}
where we have introduced the $SV$ two-point function defined as 
\begin{equation}
T_{\beta }^{SV}\left( k_{1},k_{2};m\right) =\int \frac{d^{4}k}{(2\pi )^{4}}%
Tr\left\{ \widehat{1}\frac{1}{(\not{k}+{\not{k}}_{1})-m}\gamma _{\beta }%
\frac{1}{(\not{k}+{\not{k}}_{2})-m}\right\} .
\end{equation}
The eq.(17) can be stated before the introduction of the integration sign.

At this point we can ask ourselves for the meaning of the eqs.(12), (15),
(16) and (17) and why they are important for our present investigation.
First, we note that all the involved mathematical structures are, in
principle, divergent quantities. This means that, in order to specify in a
definite way the corresponding physical amplitudes, it will be necessary to
handle undefined mathematical quantities. This implies in taking choices for
the arbitrariness involved. Since we cannot run away from some assumptions,
the unique guides we have at our disposal are the physical constraints we
can eventually identify. The eqs.(12), (15) and (16) work like constraints
for the explicit calculations, i.e., when we evaluate the $AVV$ amplitude
and after this contract the obtained expression, it must be possible to
identify mathematical structures identical to those obtained in the
evaluation of the $AV$ and $PVV$ functions previously calculated by the same
methods. The importance of the identities resides in the fact that through
such relations we can submit the decisions about the involved arbitrariness,
in the evaluation of the $AVV$ amplitude, from which the CS term should be
extracted, to the physical constraints imposed to the $AV$ and $SV$
two-point functions. Due to the eqs.(12), (15) and (16), such structures are
expected to be identified in the evaluation of the $AVV$ amplitude. This
aspect is crucial for the controversy about the value for the Chern-Simons
term in the Extended QED. In order to show what we have announced, we first
note that if we evaluate the traces involved in the $AVV$ structure, the
answer can be written in the form \cite{Orimar2}\cite{Orimar1} 
\begin{equation}
t_{\lambda \mu \nu }^{AVV}=-4\left\{ -f_{\lambda \mu \nu }+n_{\lambda \mu
\nu }+m_{\lambda \mu \nu }+p_{\lambda \mu \nu }\right\} ,
\end{equation}
where, after the integration, only $n_{\lambda \mu \nu }$ will acquire a
linear divergence's degree. It is explicitly given by 
\begin{equation}
n_{\lambda \mu \nu }=\varepsilon _{\mu \nu \lambda \alpha }\frac{%
(k+k_{2})\cdot (k+k_{3})(k+k_{1})_{\alpha }}{\left[ (k+k_{1})^{2}-m^{2}%
\right] \left[ (k+k_{2})^{2}-m^{2}\right] \left[ (k+k_{3})^{2}-m^{2}\right] }%
,
\end{equation}
which can be conveniently reorganized as 
\begin{eqnarray}
n_{\lambda \mu \nu } &=&\frac{\varepsilon _{\mu \nu \lambda \alpha }}{4}%
\left\{ \frac{2k_{\alpha }+(k_{1}+k_{2})_{\alpha }}{\left[
(k+k_{1})^{2}-m^{2}\right] \left[ (k+k_{2})^{2}-m^{2}\right] }+\frac{%
2k_{\alpha }+(k_{1}+k_{3})_{\alpha }}{\left[ (k+k_{1})^{2}-m^{2}\right] %
\left[ (k+k_{3})^{2}-m^{2}\right] }\right\}  \nonumber \\
&&+\frac{\varepsilon _{\mu \nu \lambda \alpha }}{4}\left\{ \frac{%
(k_{1}-k_{2})_{\alpha }}{\left[ (k+k_{1})^{2}-m^{2}\right] \left[
(k+k_{2})^{2}-m^{2}\right] }+\frac{(k_{1}-k_{3})_{\alpha }}{\left[
(k+k_{1})^{2}-m^{2}\right] \left[ (k+k_{3})^{2}-m^{2}\right] }\right. 
\nonumber \\
&&\;\;\;\;\;\;\ \;\;\;\left. +[2m^{2}-(k_{2}-k_{3})^{2}]\frac{%
2(k+k_{1})_{\alpha }}{\left[ (k+k_{1})^{2}-m^{2}\right] \left[
(k+k_{2})^{2}-m^{2}\right] \left[ (k+k_{3})^{2}-m^{2}\right] }\right\} .
\end{eqnarray}
The first two terms contain now all the linear divergence and the ambiguous
combination of the arbitrary internal lines momenta. Given the identity (17)
it is expected that such terms are related to $SV$ two-point functions. In
fact, it is easy to verify that 
\begin{equation}
4m\frac{2k_{\alpha }+(k_{i}+k_{j})_{\alpha }}{\left[ (k+k_{i})^{2}-m^{2}%
\right] \left[ (k+k_{j})^{2}-m^{2}\right] }=tr\left[ \widehat{1}\frac{1}{(%
\not{k}+{\not{k}}_{i})-m}\gamma _{\alpha }\frac{1}{(\not{k}+{\not{k}}_{j})-m}%
\right] .
\end{equation}
After the integration in the momentum $k,$ the right hand side can be
identified with the $SV$ two-point function defined in eq.(18). The
important aspect involved resides in the fact that all the undefined pieces
present in the $AVV$ amplitude are linked with the value of the $SV$
physical amplitude. Consequently, we can make use of the eventual physical
constraints, to be imposed on the $SV$ amplitude, to guide us in taking the
consistent choices for the corresponding arbitrariness present in the $AVV$
amplitude.

After these important remarks, in order to give additional steps to our
investigations, some manipulations and calculations involving divergent
amplitudes are required. This means to specify some strategy to handle the
problem. We adopt the calculational strategy introduced by one of us \cite
{Orimar2}\cite{Orimar3}\cite{Orimar4}\cite{Orimar1} in order to specify the
Feynman integrals which are necessary for the evaluations of all the Green's
functions involved in the present discussion.

First, we consider the two-point structures defined by 
\begin{equation}
\left( I_{2};I_{2}^{\mu }\right) =\int \frac{d^{4}k}{(2\pi )^{4}}\frac{%
\left( 1;k^{\mu }\right) }{[(k+k_{1})^{2}-m^{2}][(k+k_{2})^{2}-m^{2}]},
\end{equation}
which are given by 
\begin{eqnarray}
\bullet I_{2} &=&I_{log}(m^{2})-\left( \frac{i}{(4\pi )^{2}}\right)
Z_{0}((k_{1}-k_{2})^{2};m^{2}) \\
\bullet \left( I_{2}\right) _{\mu } &=&-\frac{1}{2}(k_{1}+k_{2})_{\alpha
}\Delta _{\alpha \mu }-\frac{1}{2}(k_{1}+k_{2})_{\mu }\left( I_{2}\right) ,
\end{eqnarray}
where we have introduced, in a more compact notation, the two-point function
structures \cite{Orimar2} 
\begin{equation}
Z_{k}(\lambda _{1}^{2},\lambda _{2}^{2},q^{2};\lambda
^{2})=\int_{0}^{1}dzz^{k}ln\left( \frac{q^{2}z(1-z)+\left( \lambda
_{1}^{2}-\lambda _{2}^{2}\right) z-\lambda _{1}^{2}}{-\lambda ^{2}}\right) ,
\end{equation}
and the basic divergent objects 
\begin{eqnarray}
\bullet \Delta _{\mu \nu } &=&\int_{\Lambda }\frac{d^{4}k}{\left( 2\pi
\right) ^{4}}\frac{4k_{\mu }k_{\nu }}{\left( k^{2}-m^{2}\right) ^{3}}%
-\int_{\Lambda }\frac{d^{4}k}{\left( 2\pi \right) ^{4}}\frac{g_{\mu \nu }}{%
\left( k^{2}-m^{2}\right) ^{2}} \\
\bullet I_{log}(m^{2}) &=&\int_{\Lambda }\frac{d^{4}k}{\left( 2\pi \right)
^{4}}\frac{1}{\left( k^{2}-m^{2}\right) ^{2}}.
\end{eqnarray}
According to the same prescription we can also calculate the integrals 
\begin{equation}
\left( I_{3};I_{3}^{\mu };I_{3}^{\mu \nu }\right) =\int \frac{d^{4}k}{(2\pi
)^{4}}\frac{\left( 1;k^{\mu };k^{\mu }k^{\nu }\right) }{\left[
(k+k_{1})^{2}-m^{2}\right] \left[ (k+k_{2})^{2}-m^{2}\right] \left[
(k+k_{3})^{2}-m^{2}\right] }.
\end{equation}
We write the results as 
\begin{eqnarray}
&&\bullet I_{3}=\left( \frac{i}{(4\pi )^{2}}\right) \xi _{00} \\
&&\bullet \left( I_{3}\right) _{\mu }=\left( \frac{i}{(4\pi )^{2}}\right)
\{\left( k_{1}-k_{2}\right) _{\mu }\xi _{01}-\left( k_{3}-k_{1}\right) _{\mu
}\xi _{10}\}-k_{1\mu }I_{3} \\
&&\bullet \left( I_{3}\right) _{\mu \nu }=\left( \frac{i}{(4\pi )^{2}}%
\right) \left\{ -\frac{g_{\mu \nu }}{2}\left[ \eta _{00}\right] \;+\left(
k_{1}-k_{2}\right) _{\mu }\left( k_{1}-k_{2}\right) _{\nu }\xi _{02}+\left(
k_{3}-k_{1}\right) _{\mu }\left( k_{3}-k_{1}\right) _{\nu }\xi _{20}\right. 
\nonumber \\
&&\;\;\;\;\;\;\left. -\left( k_{1}-k_{2}\right) _{\mu }\left(
k_{3}-k_{1}\right) _{\nu }\xi _{11}-\left( k_{1}-k_{2}\right) _{\nu }\left(
k_{3}-k_{1}\right) _{\mu }\xi _{11}\right\}  \nonumber \\
&&\;\;\;\;\;\;+\frac{g_{\mu \nu }}{4}\left[ I_{\log }\left( m^{2}\right) %
\right] +\frac{\Delta _{\mu \nu }}{4}-k_{1\mu }\left( I_{3}\right) _{\nu
}-k_{1\nu }\left( I_{3}\right) _{\mu }+k_{1\nu }k_{1\mu }I_{3}.
\end{eqnarray}
Here we have introduced the three-point function structures ${\xi }_{nm}$
defined as 
\begin{equation}
\xi _{nm}(k_{1}-k_{2},k_{3}-k_{1};m)=\int_{0}^{1}\,dz\int_{0}^{1-z}\,dy{%
\frac{z^{n}y^{m}}{Q(y,z)}},
\end{equation}
where $Q(y,z)=\left( k_{1}-k_{2}\right) ^{2}y(1-y)+\left( k_{3}-k_{1}\right)
^{2}z(1-z)+2\left( k_{1}-k_{2}\right) \cdot \left( k_{3}-k_{1}\right)
yz-m^{2},$ and 
\begin{equation}
\eta _{00}={\frac{1}{2}}Z_{0}((k_{3}-k_{2})^{2};m^{2})-\left( {\frac{1}{2}}%
+m^{2}\xi _{00})\right) +{\frac{1}{2}}\left( k_{3}-k_{1}\right) ^{2}\xi
_{10}+{\frac{1}{2}}\left( k_{1}-k_{2}\right) ^{2}\xi _{01}.
\end{equation}
This systematization is sufficient for the present discussions. The main
point is to avoid the explicit evaluation of such divergent structures, in
which case a regulating distribution needs to be specified.

It is important, at this point, to emphasize the general aspects of the
method. No shifts have been performed and, in fact, no divergent integrals
have been calculated. All the final results produced by this approach can be
mapped into those of any specific technique. The finite parts are the same
as they should be by physical reasons. The divergent parts can be easily
obtained . All we need is to evaluate the remaining divergent structures. By
virtue of this general character, the present strategy can be simply used to
systematize the procedures, even if one wants to use traditional techniques.
Those parts that depend on the specific regularization method are naturally
separated allowing us to analyze such dependence in a particular problem.
Let us now use the above results to calculate the physical amplitudes.

Substituting the values for the Feynman integrals in the corresponding
expressions for the $AV$ and $SV$ two-point functions, eq.(14) and eq.(18),
we get 
\begin{eqnarray}
\bullet T_{\mu }^{VS}(k_{1},k_{2};m) &=&(-)4m(k_{1}+k_{2})_{\beta
}[\triangle _{\beta \mu }] \\
\bullet T_{\mu \nu }^{AV}(k_{1},k_{2};m) &=&2\varepsilon _{\mu \nu \alpha
\beta }(k_{2}-k_{1})_{\beta }(k_{1}+k_{2})_{\xi }\triangle _{\xi \alpha }.
\end{eqnarray}
Note that the relation (17) is preserved by the performed calculations. We
focus on the fact that, in spite of the potentially ambiguous character of
the $AV$ and $SV$ functions, the identity (17) relating them is a
non-ambiguous one. The above expressions are the most general ones for both
mathematical structures. All the intrinsic arbitrariness are still present
in the result. They are the undefined mathematical structure $\triangle
_{\mu \nu }$ and the ambiguous combination of the internal momenta $%
k_{i}+k_{j}.$ In order to give a definite result for the physical
amplitudes, the arbitrariness should be removed through choices, which must
be made preserving the physical requirements. So, we can ask if there are
general aspects of QFT or symmetry determinations constraining the values
for the $SV$ and $AV$ structures. In fact, it is easy to see that both
two-point functions can be constrained to the identically zero value by
general arguments. First, we note that, due to unitarity, if a two-point
function like those considered is non-zero, it must have an imaginary part
at the threshold $(k_{1}-k_{2})^{2}=4m^{2}$ (Cutkosky's rules). The
remaining arbitrariness involved in the expressions (35) and (36) cannot
introduce such content to the $SV$ and $AV$ two-point functions. So, if a
non-zero value for the $SV$ and $AV$ amplitudes is assumed, as a consequence
of some choices, the unitarity is violated in both cases. In addition, if a
non-zero value is taken, the Lorentz invariance in the $SV$ amplitude and
the CPT symmetry in the $AV$ amplitude are broken. A connection between a
scalar and a vector particle is stated by the $SV$ two-point function, and a
connection between an axial and a vector particle through the $AV$ two-point
function. Another argument comes from the Ward identities analysis. The
contraction with the external momentum $k_{1}-k_{2}$ must lead us to a
definite zero value for vector Lorentz indexes, and to a proportionality
between the axial and the pseudo-scalar one in the case of the axial-vector
Lorentz indexes. There is no consistent interpretation apart from the zero
value for the $AV$ amplitude since both contractions lead us to a vanishing
value. In order to get the physically consistent result, we have at our
disposal two options: to choose the internal lines momenta such that $%
k_{1}+k_{2}=0$ or to select a regularization by the constraint $\triangle
_{\mu \nu }^{reg}=0$. We note that due to the identity (17) the values of
the two-point functions $SV$ and $AV$ are intimately related. From the point
of view of the Dimensional Regularization (DR), only the $SV$ one can be
treated and the result is identically vanishing. Due to the presence of the $%
\gamma _{5}$ Dirac matrix, or the totally antisymmetric tensor $\varepsilon
_{\mu \nu \alpha \beta }$, such a treatment cannot be performed in the $AV$
amplitude. The strategy we have adopted can be equally applied to both cases
and shows us that it is not reasonable to make choices that lead us to a
zero value for one of them and to a non-zero for the other one.

Given this argumentation we can immediately identify in these structures,
which are contained in the $AVV$ amplitude, the source of the Lorentz and
CPT symmetry breaking in the evaluation of the CS term. Any choices for the
ambiguities present in the $AVV$ structure \cite{Jackiw2}\cite{Elias}, which
imply in the attribution of a nonzero value for these contributions, will,
in fact, generate Lorentz and CPT violations, because such choices produce
non-vanishing $AV$ and $SV$ structures. So, the corresponding contributions
for the CS term cannot be considered as an implication of the QED Extended
theory but as a consequence of the adoption of an interpretation for the
arbitrariness involved, which is clearly not consistent. The importance of
this conclusion can be viewed if we evaluate $\Pi _{\lambda \mu \nu }\left(
p\right) $ using the most general expression for the $AVV$ mathematical
structure\cite{Orimar2}\cite{Orimar5} 
\begin{equation}
\Pi _{\lambda \mu \nu }\left( p\right) =\left( \frac{1}{2\pi ^{2}}\right)
\varepsilon _{\mu \nu \lambda \beta }p_{\beta }+2i\left\{ \varepsilon _{\mu
\nu \beta \sigma }p_{\beta }\Delta _{\lambda \sigma }-\varepsilon _{\mu \nu
\lambda \beta }p_{\sigma }\Delta _{\beta \sigma }\right\} .
\end{equation}
In order to be consistent with the above discussion, we must choose $%
\triangle _{\mu \nu }^{reg}=0$ eliminating then the contribution coming from
the ambiguous terms. So, given this fact, can we conclude that the
contribution for the CS term coming from $\Pi _{\lambda \mu \nu }\left(
p\right) $ is the (non-ambiguous) value $\left( \frac{1}{2\pi ^{2}}\right)
\varepsilon _{\mu \nu \lambda \beta }p^{\beta }$? Not yet! We must show that
the expression used to extract the above equation for $\Pi _{\lambda \mu \nu
}\left( p\right) $ is in agreement with the symmetry content of the QED
extended theory. In particular, the U(1) gauge symmetry was not assumed
broken in the construction of the extended theory. Another important aspect
is that concerning the low-energy limits. It is well-known that the $AVV$
amplitude should obey the soft limit: \ $\lim_{q_{\lambda }\rightarrow
0}q^{\lambda }T_{\lambda \mu \nu }^{AVV}=0.$ The most minimum consistency
requirement would force us to put any calculation of the $AVV$ structure in
accordance with these very general symmetry aspects. Otherwise, an eventual
value for the CS term again could not be considered as a consequence of the
extended theory but of the violation of other fundamental symmetries in the
intermediate steps of the calculations. It is a simple matter to check that,
by taking the explicit expression for the $AVV$ \ amplitude and by
contracting with the external momenta, we obtain \cite{Orimar2} 
\begin{eqnarray}
\bullet \left( k_{3}-k_{1}\right) _{\mu }T_{\lambda \mu \nu }^{AVV}
&=&\left( \frac{i}{8\pi ^{2}}\right) \varepsilon _{\nu \beta \lambda \xi
}\left( k_{1}-k_{2}\right) _{\beta }\left( k_{3}-k_{1}\right) _{\xi } \\
\bullet \left( k_{1}-k_{2}\right) _{\nu }T_{\lambda \mu \nu }^{AVV}
&=&-\left( \frac{i}{8\pi ^{2}}\right) \varepsilon _{\mu \beta \lambda \xi
}\left( k_{1}-k_{2}\right) _{\beta }\left( k_{3}-k_{1}\right) _{\xi } \\
\bullet \left( k_{3}-k_{2}\right) _{\lambda }T_{\lambda \mu \nu }^{AVV}
&=&-\left( \frac{i}{4\pi ^{2}}\right) \varepsilon _{\mu \nu \alpha \beta
}\left( k_{3}-k_{1}\right) _{\alpha }\left( k_{1}-k_{2}\right) _{\beta }%
\left[ 2m^{2}\xi _{00}\right] ,
\end{eqnarray}
where $\xi _{00}(\left( k_{3}-k_{2}\right) ^{2}=0)=\frac{-1}{2m^{2}}$. We
can identify the expression on the right hand side of the eq.(40) as the
calculated $PVV$ amplitude. Both ingredients above mentioned are absent from
the $AVV$ amplitude. This is not surprising because it is the same situation
we find in the Sutherland-Veltman paradox \cite{Sutherland-Veltman},
connected with the pion decay phenomenology.

The above equation represents a manifestation of a fundamental phenomenon:
the $AVV$ triangle anomaly. Note that it involves a different type of
arbitrariness in the perturbative calculations. It is not associated to the
ones related to divergences aspects. The terms which violate the U(1) gauge
symmetry come from finite contributions and therefore are not affected by an
eventual regularization scheme. The $AVV$ (symmetrized) physical amplitude
must be constructed (in an arbitrary way) by subtracting the violating term 
\begin{equation}
\left( T_{\lambda \mu \nu }^{AVV}(\left( k_{3}-k_{1}\right) ,\left(
k_{1}-k_{2}\right) )\right) _{phys}=T_{\lambda \mu \nu }^{AVV}(\left(
k_{3}-k_{1}\right) ,\left( k_{1}-k_{2}\right) )-T_{\lambda \mu \nu
}^{AVV}\left( 0\right) ,
\end{equation}
where 
\begin{equation}
T_{\lambda \mu \nu }^{AVV}\left( 0\right) =-\left( \frac{i}{8\pi ^{2}}%
\right) \varepsilon _{\mu \nu \lambda \beta }\left[ \left(
k_{3}-k_{1}\right) _{\beta }-\left( k_{1}-k_{2}\right) _{\beta }\right] .
\end{equation}
The resulting amplitude preserves the U(1) gauge symmetry and it is in
agreement with the low-energy theorem, in spite of violating the axial Ward
identity involved. This procedure is precisely the one followed in the
construction of the renormalizability of the Standard Model. So, again the
consistency in the perturbative calculations requires the same
interpretation for the same Green's function. This means to adopt for the $%
AVV$ amplitude the expression 
\begin{eqnarray}
\left( T_{\lambda \mu \nu }^{AVV}\right) _{Phys} &=&\left( \frac{i}{\left(
4\pi \right) ^{2}}\right) \left( -4\right) \left( k_{3}-k_{1}\right) _{\xi
}\left( k_{2}-k_{1}\right) _{\beta }\left\{ \varepsilon _{\nu \lambda \beta
\xi }[\left( k_{3}-k_{1}\right) _{\mu }\left( \xi _{20}+\xi _{11}-\xi
_{10}\right) \right.  \nonumber \\
&&\hspace{1in}\hspace{1in}\hspace{0.4in}\;+\left( k_{2}-k_{1}\right) _{\mu
}\left( \xi _{11}+\xi _{02}-\xi _{01}\right) ]  \nonumber \\
&&\hspace{1in}\hspace{1in}\;\;+\varepsilon _{\mu \lambda \beta \xi }[\left(
k_{3}-k_{1}\right) _{\nu }\left( \xi _{11}+\xi _{20}-\xi _{10}\right) 
\nonumber \\
&&\hspace{1in}\hspace{1in}\hspace{0.4in}\;+\left( k_{2}-k_{1}\right) _{\nu
}\left( \xi _{02}+\xi _{11}-\xi _{01}\right) ]  \nonumber \\
&&\hspace{1in}\hspace{1in}\;\;+\varepsilon _{\mu \nu \beta \xi }[\left(
k_{3}-k_{1}\right) _{\lambda }\left( \xi _{11}-\xi _{20}+\xi _{10}\right) 
\nonumber \\
&&\hspace{1in}\hspace{1in}\hspace{0.4in}\;-\left. \left( k_{2}-k_{1}\right)
_{\lambda }\left( \xi _{02}-\xi _{01}-\xi _{11}\right) ]\right\}  \nonumber
\\
&&-\left( \frac{i}{\left( 4\pi \right) ^{2}}\right) \varepsilon _{\mu \nu
\lambda \beta }\left( k_{3}-k_{1}\right) _{\beta }\left\{ Z_{0}\left( \left(
k_{1}-k_{3}\right) ^{2};m^{2}\right) -Z_{0}\left( \left( k_{2}-k_{3}\right)
^{2};m^{2}\right) \right.  \nonumber \\
&&\hspace{1in}\hspace{1in}+\left[ 2\left( k_{3}-k_{2}\right) ^{2}-\left(
k_{1}-k_{3}\right) ^{2}\right] \xi _{10}+  \nonumber \\
&&\hspace{1in}\hspace{1in}\left. -\left( k_{1}-k_{2}\right) ^{2}\xi _{01}+%
\left[ 1-2m^{2}\xi _{00}\right] \right\}  \nonumber \\
&&-\left( \frac{i}{\left( 4\pi \right) ^{2}}\right) \varepsilon _{\mu \nu
\lambda \beta }\left( k_{2}-k_{1}\right) _{\beta }\left\{ Z_{0}\left( \left(
k_{1}-k_{2}\right) ^{2};m^{2}\right) -Z_{0}\left( \left( k_{2}-k_{3}\right)
^{2};m^{2}\right) \right.  \nonumber \\
&&\hspace{1in}\hspace{1in}+\left[ 2\left( k_{3}-k_{2}\right) ^{2}-\left(
k_{1}-k_{2}\right) ^{2}\right] \xi _{01}  \nonumber \\
&&\hspace{1in}\hspace{1in}\left. -\left( k_{3}-k_{1}\right) ^{2}\xi _{10}+%
\left[ 1-2m^{2}\xi _{00}\right] \right\} -T_{\lambda \mu \nu }^{AVV}\left(
0\right) .
\end{eqnarray}
Now, taking the kinematical situation where the CS term is defined, eq.(9),
we get 
\begin{equation}
\Pi _{\mu \nu \lambda }(p)=\left( \frac{1}{2\pi ^{2}}\right) \varepsilon
_{\mu \nu \lambda \beta }p_{\beta }-iT_{\lambda \mu \nu }^{A\rightarrow
VV}\left( 0\right) .
\end{equation}
Identifying then $T_{\lambda \mu \nu }^{AVV}\left( 0\right) $ as the
violating term on the left hand side of the eq.(38) and (39) this means that
the identically vanishing value is obtained.

So, a clean and sound conclusion is extracted: the consistency in
perturbative calculations leave no room for the existence of the radiatively
induced CS term in the extended QED. Therefore, if one wants to get a
nonzero value for such contribution, it is necessary: {\bf 1)} to break in
the intermediary steps of the calculation, Lorentz, CPT, unitarity and an
axial Ward identity by attributing to mathematical structures, identical to
the two-points functions $AV$ and $SV$, which are related, a nonzero value
or, {\bf 2)} to violate the low-energy theorem $\lim_{q_{\lambda
}\rightarrow 0}q^{\lambda }T_{\lambda \mu \nu }^{AVV}=0,$ which may imply
simultaneously in the violation of U(1) gauge symmetry in the Extended QED.
The implication of the last sentence is the spoiling of the Standard Model
renormalizability by destroying the anomaly cancellation mechanism. Any of
such options clearly implies in ignoring the wider sense of the consistency
in perturbative calculations, which means to treat the same Green's function
in the same way in all places where they occur. If one does not consider
these aspects, in fact, one can obtain Lorentz and CPT violation not only
for the discussed problem but, following the same recipe, it is possible to
state a copious number of similar situations in other theories and models.

{\bf Acknowledgements}: G.D. acknowledge a grant from CNPq/Brazil and O.A.B.
from FAPERGS/Brazil.

\end{document}